\documentclass[review]{elsarticle}

\usepackage{amssymb}
\usepackage{nomencl}
\usepackage{amsmath}
\usepackage{amsfonts}
\usepackage{ragged2e}
\usepackage{graphicx}
\usepackage{txfonts}
\usepackage{booktabs}
\usepackage{tikz}
\usepackage{palatino}
\usepackage{mathpazo}
\usepackage[mathscr]{euscript}
\usepackage{mathalpha}
\usepackage{notoccite}
\usepackage{multirow}
\usepackage{subfiles,tocbibind,array,url,pdflscape,nomencl,fancyhdr,etoolbox}

\makenomenclature

\newcommand{\cbulkinitial}{\ensuremath{{C_{\mathrm{bulk,0}}}}}

\newcommand{\cbulk}{\ensuremath{{C_{\mathrm{bulk}}}}}
\newcommand{\cs}{\ensuremath{{C_{\mathrm{s}}}}}
\newcommand{\cparticle}{\ensuremath{{C_{\mathrm{p}}}}}
\newcommand{\csinterface}{\ensuremath{{C_{\mathrm{s,interface}}}}}

\newcommand{\xb}{\ensuremath{{X_{\mathrm{b}}}}}

\newcommand{\ds}{\ensuremath{\mathscr{D}_\mathrm{s}}}
\newcommand{\dsparticle}{\ensuremath{\mathscr{D}_\mathrm{p}}}
\newcommand{\dbulk}{\ensuremath{\mathscr{D}_\mathrm{bulk}}}
\newcommand{\yb}{\ensuremath{{Y_{\mathrm{b}}}}}
\newcommand{\khalfsat}{\ensuremath{{k_{\mathrm{2}}}}}
\newcommand{\kbiomassinactive}{\ensuremath{{k_{\mathrm{3}}}}}
\newcommand{\kparticle}{\ensuremath{K_\mathrm{p}}}
\newcommand{\kf}{\ensuremath{k_\mathrm{f}}}

\newcommand{\kfp}{\ensuremath{k_\mathrm{f,p}}}
\newcommand{\Shp}{\ensuremath{{Sh}_\mathrm{p}}}

\newcommand{\ebulk}{\ensuremath{{\epsilon_{\mathrm{l}}}}}
\newcommand{\epart}{\ensuremath{{\epsilon_{\mathrm{geo}}}}}
\newcommand{\fpack}{\ensuremath{{\epsilon_{\mathrm{c}}}}}
\newcommand{\xparticlehat}{\ensuremath{\hat{{x}}}}
\newcommand{\freqpart}{\ensuremath{\nu_\mathrm{p}}}
\newcommand{\nparticle}{\ensuremath{\phi_\mathrm{p}}}

\newcommand{\surfpart}{\ensuremath{S_\mathrm{p}}}

\newcommand{\dparticle}{\ensuremath{{d_{\mathrm{cell}}}}}
\newcommand{\constone}{\ensuremath{P_\mathrm{1}}}
\newcommand{\consttwo}{\ensuremath{P_\mathrm{2}}}
\newcommand{\constthree}{\ensuremath{P_\mathrm{3}}}
\newcommand{\constfour}{\ensuremath{P_\mathrm{4}}}
\newcommand{\rhoparticle}{\ensuremath{{\rho_{\mathrm{p}}}}}
\newcommand{\ybparticle}{\ensuremath{{Y_{\mathrm{p}}}}}
\newcommand{\etaparticle}{\ensuremath{{\eta}}}
\newcommand{\rhoxb}{\ensuremath{{\rho_{\mathrm{b}}}}}
\newcommand{\rhoedr}{\ensuremath{\hat{{\rho}}}}

\newcommand{\ninitialparticle}{\ensuremath{{{\phi^{\mathrm{o}}_\mathrm{geo}}}}}

\newcommand{\aspbulk}{\ensuremath{{A_{\mathrm{sp,bulk}}}}}
\newcommand{\Abiofilm}{\ensuremath{{A_{\mathrm{s}}}}}

\newcommand{\lf}{\ensuremath{L_\mathrm{f}}}

\newcommand{\lpart}{\ensuremath{L_\mathrm{p}}}
\newcommand{\Wchannelradius}{\ensuremath{{W_{\mathrm{x}}}}}
\newcommand{\Wchannel}{\ensuremath{{W_{\mathrm{system}}}}}
\newcommand{\Hchannel}{\ensuremath{{H_{\mathrm{system}}}}}

\newcommand{\Vbiofilmparticleeff}{\ensuremath{{V_{\mathrm{p,eff}}}}}

\nomenclature{$\cbulkinitial$}{Initial oxygen concentration in bulk phase, mg oxygen L$^{-1}$ bulk volume}
\nomenclature{$\cbulk$}{Oxygen concentration in bulk phase, mg oxygen L$^{-1}$ bulk volume}
\nomenclature{$\cs$}{Oxygen concentration in biofilm phase,mg oxygen L$^{-1}$ biofilm volume}
\nomenclature{$\cparticle$}{Oxygen concentration in capsule, mg oxygen L$^{-1}$ biofilm volume}
\nomenclature{$\csinterface$}{Oxygen concentration at bulk-biofilm interface, mg oxygen L$^{-1}$ biofilm volume}
\nomenclature{$\xb$}{Microbial biomass concentration in biofilm phase, mg biomass L$^{-1}$ biofilm volume}
\nomenclature{$z$}{Lateral dimension, m}
\nomenclature{$x$}{Radial dimension, m}
\nomenclature{$\ds$}{Effective diffusion coefficient of oxygen in biofilm phase, m$^{2}$ day$^{-1}$}
\nomenclature{$\dsparticle$}{Effective diffusion coefficient of oxygen in capsule, m$^{2}$ day$^{-1}$}
\nomenclature{$\dbulk$}{Effective diffusion coefficient of oxygen in bulk phase, m$^{2}$ day$^{-1}$}
\nomenclature{$\yb$}{Yield coefficient for oxygen utilization, g biomass g$^{-1}$ oxygen}
\nomenclature{$\khalfsat$}{Half-maximum rate concentration of oxygen, mg oxygen L$^{-1}$ biofilm volume}
\nomenclature{$\kbiomassinactive$}{Rate decay coefficient for cell-capsule structures relative to changes in EPS matrix, day$^{-1}$}
\nomenclature{$\kparticle$}{Dimensional rate constant for oxygen utilization in capsule, day$^{-1}$}
\nomenclature{$\kf$}{Oxygen mass transfer coefficient at bulk-biofilm interface, m s$^{-1}$}
\nomenclature{$\kfp$}{Rate coefficient at biofilm-capsule interface, day$^{-1}$}
\nomenclature{$\Shp$}{Sherwood number at biofilm-capsule interface, dimensionless}
\nomenclature{$\ebulk$}{Bulk volume fraction, dimensionless}
\nomenclature{$\epart$}{Biofilm volume fraction, dimensionless}
\nomenclature{$\fpack$}{Compactness factor, dimensionless}
\nomenclature{$\xparticlehat$}{Distribution mean in terms of number of cell-capsule structures or its count associated with the event under consideration per unit biofilm surface area, dimensionless}
\nomenclature{$\freqpart$}{Effective maximum growth rate for oxygen utilization based on cell-capsule patterns, number of cell-capsule structures per m$^{2}$ biofilm surface area per day}
\nomenclature{$\nparticle$}{Apparent number density based on cell-capsule patterns, and expressed in terms of number of cell-capsule structures per unit biofilm surface area, dimensionless}
\nomenclature{$\surfpart$}{Surface area of the particle, m$^{2}$}
\nomenclature{$\dparticle$}{Characteristic diameter of the cell, m}
\nomenclature{$\delta$}{Shape factor, dimensionless}
\nomenclature{$\sigma$}{Scale factor, dimensionless}
\nomenclature{$\constone$}{First parameter based on shape factor and scale factor, dimensionless}
\nomenclature{$\consttwo$}{Second parameter based on distribution mean in terms of number of cell-capsule structures or its count associated with the event under consideration, dimensionless}
\nomenclature{$\constthree$}{Third parameter based on scale factor, dimensionless}
\nomenclature{$\constfour$}{Fourth parameter based on shape factor, dimensionless}
\nomenclature{$\rhoparticle$}{Capsule density, mg biomass L$^{-1}$ biofilm volume}
\nomenclature{$\ybparticle$}{True-yield of EPS produced for oxygen utilization, g EPS g$^{-1}$ oxygen}
\nomenclature{$\rhoxb$}{Biomass density, mg biomass L$^{-1}$ biofilm volume}
\nomenclature{$\rhoedr$}{Normalized density, dimensionless}
\nomenclature{$\ninitialparticle$}{Initial number of cell-capsule structures per unit biofilm surface area at the start of biofilm development, dimensionless}
\nomenclature{$\lf$}{Biofilm thickness, m}
\nomenclature{$\lpart$}{Capsule thickness, m}
\nomenclature{$\Wchannelradius$}{Half-width of system domain, m}
\nomenclature{$\Wchannel$}{Width of system domain, m}
\nomenclature{$\Hchannel$}{Height of system domain, m}
\nomenclature{$\Vbiofilmparticleeff$}{Effective volume of cell-capsule structures, m$^{3}$}

\usepackage{lineno,hyperref}

\journal{arXiv}

\biboptions{authoryear}

\begin{document}
	
	\begin{frontmatter}
		
		\title{Modeling multiscale architecture of biofilm extracellular matrix and its role in oxygen transport}
		
		\author[rvt]{Raghu K. Moorthy}
		\author[rvt]{Eoin Casey \corref{mycorrespondingauthor}}
		\ead{eoin.casey@ucd.ie}
		\address[rvt]{School of Chemical and Bioprocess Engineering, University College Dublin, Belfield, Dublin, Ireland}
		
		\cortext[mycorrespondingauthor]{Corresponding author}

		
		\begin{abstract}
		The extracellular polymeric substances (EPS) matrix of microbial biofilms exhibits a complex structural heterogeneity that profoundly influences mass transport and metabolic activity. Conventional biofilm models typically assume a homogeneous matrix, thereby neglecting the localized transport resistance introduced by the bacterial capsule, a distinct, low-diffusivity polysaccharide layer surrounding individual cells. In this theoretical study, we develop a multiscale "cell–capsule" continuum model that represents the capsule as a concentric shell enveloping each microbial cell core within the bulk EPS. Utilizing a one-dimensional reaction–diffusion framework coupled with a geometric characterization of capsule spacing and thickness, we quantify how microscale architecture modulates oxygen transport in developing biofilms. Model simulations demonstrate that incorporating a discrete capsular phase introduces a pronounced "resistance-in-series" effect, reducing local oxygen availability by up to 70\% compared to conventional homogeneous models. Furthermore, our analysis indicates that capsule thickness and matrix compaction jointly control the effective diffusivity and oxygen effectiveness factor within the biofilm. These results provide critical mechanistic insights into how microscale organization governs macroscale biofilm function, offering a new framework for integrating structural heterogeneity into multiscale biofilm simulations.
		\end{abstract}
		
	\end{frontmatter}
	
	\section{Introduction}
	Microbial cells encased in an extracellular matrix form multicellular structures known as biofilms (\cite{Costerton1987a}; \cite{Rittmann1980}; \cite{Wanner1995}). The biofilm matrix consists of insoluble polymers that promote heterogeneity in the biofilm (\cite{Sutherland2001}; \cite{Li2018}; \cite{Schleheck2009}; \cite{Quan2022}; \cite{Ebrahimi2019}; \cite{Pechaud2024}). A structural feature of biofilm  that is rarely incorporated in mathematical models of biofilms is the polysaccharide capsule enveloping the microbial cell wall (\cite{Eichner2025a}; \cite{Phanphak2019a}; \cite{Bayer1990a}), which may play a significant role in the biofilm phenotype (\cite{Whitfield2020a}; \cite{Gao2024a}). These polysaccharides are generated at the cell wall in response to diverse environmental conditions, suggesting a possible survival mechanism in various biofilm-forming organisms (\cite{Cleary1979}; \cite{Weiser2001}; \cite{Geno2014a}; \cite{Saren2026a}). Previous reviews have noted several key features of the capsule, including its thickness (see Table \ref{tab:1D-model-capsule-thickness-summary}) and close resemblance to a shell-like structure, originally proposed in 1978 by William Characklis \citep{Characklis1978}. However, little is known about the role of capsule in areas such as the transport of nutrients within biofilms, particularly since it is now emerging that capsule is quite different from the bulk EPS. 
	\par In the context of biofilm modeling, the conventional approach is to assume homogeneity of the matrix (\cite{Duddu2009}; \cite{Mattei2018}). Over the past forty years most mathematical models of biofilms have described the biofilm matrix as a homogenous phase, overlooking its intricate structural features (\cite{Wanner1986}; \cite{Wanner1995}; \cite{Drury1993b}). In comparisons to discrete models \citep{PritchettDockery2010b}, developing structurally complex continuum models offer distinct benefits by incorporating recent experimental advances in biofilm microstructure (\cite{Geno2014a}; \cite{Eichner2025a}; \cite{Hasan2025a}).
	State-of-the-art mathematical models for biofilm development based on individual biomass particles include the landmark paper by Lardon and co-workers (and popularly known as iDynoMiCS) \citep{Lardon2011}, which was preceded by the particle-based 2D/3D model for biofilm growth mechanisms as developed by Picioreanu and co-workers using discrete positions based on adjacent probabilities (cellular automata approach) \citep{Picioreanu2004}. The strength of these "bottom-up" approaches is that they treat each microbe as a discrete entity with individual properties, enabling the simulation of heterogeneous microenvironments. However, a major limitation is their reliance on specific, highly detailed parameters for individual cellular characteristics, for which measurements are rarely available and the detailed understanding of some physical interactions (e.g. detachment/dispersion) which restrict the model's application and accuracy.
	\par In the present study, we address these limitations by introducing a multiscale modeling approach that explicitly partitions the structural properties of the bacterial capsule from the bulk EPS matrix. We propose a novel cell–capsule framework (analogous to the core–shell approach utilized in chemical reaction engineering) to characterize the extracellular matrix and investigate localized oxygen transport within the biofilm. We hypothesize that accounting for the distinct material properties of the capsule versus the bulk EPS will yield more accurate mass transfer predictions. Using previously published empirical datasets on the poroelastic properties of both fractions, we quantify the relative differences in oxygen transport rates. We expect this framework to improve predictive models of oxygen depletion in complex biofilm microenvironments, thereby enhancing the optimization of biofilm reactor strategies \citep{Westbrook2018a} and informing the rational design of capsule-targeted antimicrobial therapies \citep{Formosa2012a}.
	
	\begin{table}[tbp]
		\caption{Landmark studies on bacterial capsule morphology}
		\label{tab:1D-model-capsule-thickness-summary}
		\centering
		\begin{tabular}{p{5cm}lll}
			\toprule
			\multirow{2}{*}{Morphology} &
			\multicolumn{2}{l}{Capsule thickness (nm)} &
			\multirow{2}{*}{Reference} \\ \cmidrule(lr){2-3}
			& Minimum & Maximum & \\ \cmidrule(lr){1-4}
			Bimodal distribution of thick capsule around the core & 50 & 800 & \cite{Bayer1990a} \\
			Surface structures and randomly organized capsule around the core & 300 & 500 & \cite{Stukalov2008a} \\ 
			Fimbriae structures, and with an organized capsule around the core & 378 & 462 & \cite{Wang2015a} \\
			Bimodal distribution of thick capsule as polymeric brushes around the core & 197 & 385 & \cite{Phanphak2019a} \\
			Random capsule thickness within the same serotype & 150 & 700 & \cite{Eichner2025a} \\ 
			\bottomrule \\
		\end{tabular}
	\end{table}
	
	\section{Methodology}
	\subsection{System description}
	The system domain consists of two coupled compartments: the biofilm phase and the bulk phase. A one-dimensional analysis suffices to represent the microbial biofilm. The substratum is inert and does not influence the kinetics governing the biofilm formation. Monod kinetics describes the substrate utilization for biomass growth at steady-state (\cite{PritchettDockery2001a}; \cite{Schleheck2009}; \cite{Mattei2018}; \cite{Wing2024}). We propose a continuum approach to conceptualize the biofilm with varying physiochemical properties, where each microbial cell (cell as core) is surrounded by capsular matrix (capsule as shell), forming a "cell-capsule" structure.
	
	\subsection{Model assumptions}
	In the biofilm phase, microbial cells and the insoluble polymers of the extracellular matrix (or EPS) collectively are known as particulate matter \citep{Quan2022}. Fick’s law of mass diffusion applies to oxygen transport in both the biofilm phase and the capsule structure. The diffusion coefficient for oxygen in the biofilm phase is assumed 0.8 times that in the bulk phase \citep{Stewart1998}. In contrast, the diffusion transport of oxygen within the capsule is assumed 0.2 times than that in the bulk phase (\cite{Cleary1979}; \cite{McCabe1975a}), based on our estimates using published data of capsule material properties (see supplementary details, Tables S1 to S4). Analysis of these poroelastic datasets indicated that for the capsule, the effective diffusion coefficient relative to the bulk phase consistently falls within the above range. Specifically, the data demonstrates a saturation threshold supported by the biofilm volume fraction, where the capsule’s internal structure becomes sufficiently dense that further increases in Young's modulus (beyond 6 kPa) do not significantly reduce the available free volume for oxygen transport \citep{Huang2026a}. Assuming a linear concentration gradient across the bulk-biofilm interface, the mass transfer coefficient for oxygen transport is estimated using a theoretical correlation for the biofilm system \citep{Comiti2000}.
	
	\subsection{Model formulation}
	\label{model formulated}
	\subsubsection{Distribution of cell-capsule structures}
	In this study, we investigate how randomly arranged, spherical structures replicate the morphology of the biofilm. Given the number of cell structures occupying per unit surface area of the biofilm in unit time ($\nparticle$), the compactness factor is denoted by $\fpack$ so as to represent the cellular arrangements or spacing. Previous research shows that upon binning these representative geometric constructs, their cumulative numbers can increase monotonically (\cite{Ebrahimi2019}; \cite{Melaugh2023}), until they level off when the maximum number of structures effectively mimics the biofilm morphology (see supplementary details, Figure S1). Under steady-state conditions, the apparent number density based on cell-capsule patterns is referred to as $\freqpart$ in our model. The key parameters required are the shape factor ($\delta$) and the scale factor ($\sigma$) of the geometry to obtain the aforementioned number density (that is, in general, $f(x)$ value as in Equation \ref{eqn:particulates-PSD-general} \citep{Habibullah2000}).
	
	\begin{equation}
		f(x) = \frac{\delta}{\sigma}{\left(1+\left(\frac{x-\xparticlehat}{\sigma}\right)\right)}^{-\left(\delta+1\right)}
		\label{eqn:particulates-PSD-general}
	\end{equation}  
	where, $x$ is number of cell-capsule structures, $f\left(x\right)$ is a density function in terms of number of cell-capsule structures per unit surface area, $\xparticlehat$, $\sigma$ and $\delta$ are the arithmetic mean, scale factor (or standard deviation) and a shape factor to represent the cell-capsule (geometrical) population. \\
	For simplicity, the above notations are put together into a set of four parameters as follows (see appendix details for the mathematical derivations):
	
	\begin{subequations}
		\begin{align}
			f(x) = \constone \left(1+\left(\frac{x+\consttwo}{\constthree}\right)\right)^{-\constfour} \\
			\text{where, }\constone = \frac{\delta}{\sigma} \\
			\consttwo = - \xparticlehat \\
			\constthree = \sigma \\
			\constfour = \delta + 1
		\end{align}
		\label{eqn:particulates-PSD-general-parameters}
	\end{subequations}

	\subsubsection{Model parameters}
	\label{BPM-parameters}
	We propose two model parameters to evaluate the cellular arrangements in the biofilm phase. These include capsule thickness ($\lpart$) and the compaction factor ($\fpack$). The cell-capsule structures are described by $\fpack$ reflecting on the compactly arranged matrix within the biofilm. Table \ref{tab:2D-model-without-flow-coupled-transport-parameters-list} summarizes a sample set of assumed values based on the initial number of cell-capsule structures ($\ninitialparticle$) and the total volume fraction ($\epart$). The $\fpack$ values are then calculated from the following correlation (Equation \ref{eqn:packing-fraction-estimated}): \\
	\begin{equation}
		\fpack = \frac{1}{\left(\frac{1}{\epart}-1\right)}
		\label{eqn:packing-fraction-estimated}
	\end{equation} \\
	\par For each combination of $\lpart$ and $\fpack$, the corresponding Sherwood number for oxygen mass transfer, $\Shp$, is estimated. By defining an effectiveness factor, $\etaparticle$, we describe the oxygen mass transfer across the biofilm-capsule interface. It is defined as the actual mass transfer rate divided by the standard rate which would be obtained with no diffusion resistance. The following correlations are used in our model to compare different cell-capsule patterns (Equations \ref{eqn:packing-fraction-estimated} to \ref{eqn:particle-effectiveness-number-formulation}):    \\
	\begin{equation}
		\Shp = \frac{\kfp \left(0.5\dparticle+\lpart\right)}{\dsparticle}
		\label{eqn:particle-Sherwood-number-formulation}
	\end{equation} \\
	where, $\dparticle$ is cell diameter, $\dsparticle$ is effective diffusion coefficient of oxygen in the capsule, and $\kfp$ is the rate coefficient at the biofilm-capsule interface approximating to a characteristic $\dsparticle$ when scaled by the cellular length scale. \\
	\begin{equation}
		\etaparticle = \left[\frac{\fpack\kparticle}{\freqpart}\right]{\left[\frac{\left(\Shp\right)_{\text{without mass transfer}}}{\Shp}\right]}	
		\label{eqn:particle-effectiveness-number-formulation}
	\end{equation} \\
	where, $\freqpart$ representing the effective maximum specific growth rate due to oxygen utilization in the biofilm is obtained from Equation \ref{eqn:Pareto-distribution} (see Appendix for more details, or section \ref{AppendixEquations}).

	\subsection{Image analysis}
	The input datasets for our model are generated based on an assumed cell diameter ($\dparticle$ $\approx$  1 $\mu$m). A fixed capsule thickness around the cell constituted each of these structures. An increase in the capsule thickness distinctly represents both capsule and the bulk EPS matrix (see supplementary details, Figure S2). The cell-capsule patterns are adjusted using specific spacing between equally sized, perfectly spherical geometry to mimic the compaction factor ($\fpack$) hypothesized from previous literature (\cite{Quan2022}; \cite{Stewart1998}). Probabilistic, randomly distributed cell-capsule structures inside a unit surface area is adapted to generate at least nine independent snapshots (MATLAB) (The MathWorks, Inc., USA) (see supplementary details for the corresponding MATLAB code, or on our project repository web page hosted at GitHub \citep{Moorthy2025a}). These images are then analyzed using suitable thresholds and 'Analyze Particles' tool within the ImageJ software (ImageJ, NIH, USA).
	
	\subsection{Numerical simulations of the model}
	\label{model implementation}
	Adapting the solution methodology as followed in standard biofilm models (\cite{Mattei2018}; \cite{Wanner1986}), the finite difference method numerically calculates the oxygen concentration in the capsule ($\cparticle$), and in the biofilm phase ($\cs$) (Table \ref{tab:2D-model-without-flow-coupled-transport-parameters-list}, and see supplementary details, Figure S3). We begin with a fixed cell diameter ($\dparticle$), a constant biomass density ($\rhoxb$), and an initial bulk phase oxygen concentration ($\cbulkinitial$) under steady-state conditions (see stoichiometric matrix for the present study, Table \ref{tab:1D-model-without-flow-coupled-transport-schematic-stoichiometry-matrix}). Using an extracted value of $\fpack$ from the above image analysis, we solve Equation \ref{eqn:particle-density-estimated} to estimate the capsule density ($\rhoparticle$). The confined system domain is simulated with a constant oxygen concentration in the bulk phase ($\cbulkinitial$), and compared to that at the bulk-biofilm interface ($\cs$), for every $z$ to obtain the spatial concentration profiles. Further details on the mathematical equations and definitions of system domain boundaries are available in the Appendix (see section \ref{AppendixEquations}).
	\par Standard biofilm models following Monod kinetics are numerically solved to simulate a benchmark problem. We compared our model with these simulations at fixed initial oxygen concentrations in the bulk phase and capsule thickness. We evaluated for any deviations from the standard model at two different maximum biomass growth rates (0.1 and 1 day$^{-1}$). Considering simplicity in the modelling approach, we assumed a constant mass transfer flux at the biofilm-capsule interface (\cite{Stewart2016}; \cite{Geno2014a}) to simulate oxygen penetration in the matrix (see Table \ref{tab:2D-model-without-flow-coupled-transport-parameters-list-benchmark-analysis} for the list of model parameters used in the benchmark analysis).
	
	\begin{subequations}
		\begin{align}
			\rhoedr = \left(\frac{\yb}{\ybparticle}\right)\left(\frac{1}{\fpack}+1\right) \\
			\rhoparticle = \rhoxb \rhoedr
		\end{align}
		\label{eqn:particle-density-estimated}
	\end{subequations}
	
	\begin{table}[tbp]
		\caption{Stoichiometric matrix for the present study}
		\label{tab:1D-model-without-flow-coupled-transport-schematic-stoichiometry-matrix}
		\centering
		\begin{tabular}{p{6cm}ll}
			\toprule
			Process & Oxygen as substrate & Rate expression \\ \midrule
			Microbial biomass growth due to Monod kinetics & -$\frac{1}{\yb}$ & {$\frac{\freqpart \cs \xb}{\khalfsat + \cs}$} \\
			EPS production due to oxygen utilization in the capsule & -$\frac{1}{\ybparticle}$ & {$\frac{\kparticle \cparticle \rhoparticle}{\khalfsat + \cparticle}$} \\
			\bottomrule \\
		\end{tabular}
	\end{table}
	
	\begin{table}[tbp]
		\caption{List of model parameters for the present study}
		\label{tab:2D-model-without-flow-coupled-transport-parameters-list}
		\centering
		\begin{tabular}{lp{7.5cm}p{2cm}p{3.5cm}}
			\toprule
			Parameter & Description & Value & Reference \\ \midrule
			$\cbulkinitial$ & Initial oxygen concentration (mg L$^{-1}$) & 6 & \cite{Wanner2006} \\
			$\ds$ & Effective diffusion coefficient of oxygen in biofilm phase (m$^{2}$ day$^{-1}$) & 4*10$^{-4}$ & \cite{Wanner2006}; \cite{Stewart1998} \\
			$\rhoxb$ & Biomass density, (mg L$^{-1}$) & 778 & \cite{Wanner2006} \\
			$\yb$ & Yield coefficient for oxygen utilization (dimensionless) & 0.25 & \cite{Wanner2006} \\
			$\khalfsat$ & Half-maximum rate concentration of oxygen (mg L$^{-1}$) & 50 & Assumed \\
			$\kbiomassinactive$ & Rate decay coefficient for cell-capsule structures relative to changes in EPS matrix (day$^{-1}$) & 10$^{3}$ & Assumed \\
			$\Wchannel$ & Breadth of rectangular domain (mm) & 3.80 & Typical commercial flow cell dimension \\
			$\Hchannel$ & Height of rectangular domain (mm) & 0.40 & Typical commercial flow cell dimension \\
			$\lf$ & Biofilm thickness (m) & 10*10$^{-6}$ & Assumed \\
			$\delta$ & Shape factor (dimensionless) & 0.50 & Assumed for spherical geometry \\
			$\sigma$ & Scale factor (dimensionless) & 10$^{-2}$ & Assumed \\
			$\ninitialparticle$ & Initial number of cell-capsule structures (dimensionless) & 10$^{3}$ & Assumed \\
			$\consttwo$ & Standard deviation for number of cell-capsule structures (dimensionless) & 2 & Assumed \\
			$\epart$ & Total volume fraction in the biofilm (dimensionless) & Varied (0.95 to 0.02) & \cite{Stewart1998}; \cite{Quan2022} \\
			$\fpack$ & Compaction factor (dimensionless) & Varied (0.05 to 50) & Estimated \\
			\bottomrule
		\end{tabular}
	\end{table} 
	
	\begin{table}[tbp]
		\caption{List of model parameters for the benchmark analysis}
		\label{tab:2D-model-without-flow-coupled-transport-parameters-list-benchmark-analysis}
		\centering
		\begin{tabular}{lp{7.5cm}p{2cm}p{3.5cm}}
			\toprule
			Parameter & Description & Value & Reference \\ \midrule
			$\cbulkinitial$ & Initial oxygen concentration (mg L$^{-1}$) & [5, 9.5] & \cite{Rittmann1980} \\
			$\ds$ & Effective diffusion coefficient of oxygen in biofilm phase (m$^{2}$ day$^{-1}$) & 0.1*10$^{-4}$ & \cite{Rittmann1980} \\
			$\rhoxb$ & Biomass density (mg L$^{-1}$) & 40000 & \cite{Rittmann1980} \\
			$\yb$ & Yield coefficient for oxygen utilization (dimensionless) & 0.5 & \cite{Rittmann1980} \\
			$\khalfsat$ & Half-maximum rate concentration of oxygen (mg L$^{-1}$) & 3.4 & \cite{Rittmann1980}; \cite{Stewart2016} \\
			$\kbiomassinactive$ & Rate decay coefficient for cell-capsule structures relative to changes in EPS matrix (day$^{-1}$) & 10$^{3}$ & Assumed \\
			$\Wchannel$ & Breadth of rectangular domain (mm) & 3.80 & Typical commercial flow cell dimension \\
			$\Hchannel$ & Height of rectangular domain (mm) & 0.40 & Typical commercial flow cell dimension \\
			$\delta$ & Shape factor (dimensionless) & 0.50 & Assumed for spherical geometry \\
			$\sigma$ & Scale factor (dimensionless) & 1 & Assumed \\
			$\ninitialparticle$ & Initial number of cell-capsule structures (dimensionless) & 10$^{5}$ & Assumed \\
			$\epart$ & Total volume fraction in the biofilm (dimensionless) & 0.15 & \cite{Stewart1998}; \cite{Quan2022} \\
			$\fpack$ & Compaction factor (dimensionless) & 6.677 & Assumed \\
			\bottomrule
		\end{tabular}
	\end{table}

	\section{Results and Discussion}
	
	\subsection{Cell-capsule structure as a 'resistance-in-series' model}
\begin{figure}[tbp]
		\includegraphics[width=1.0\textwidth]{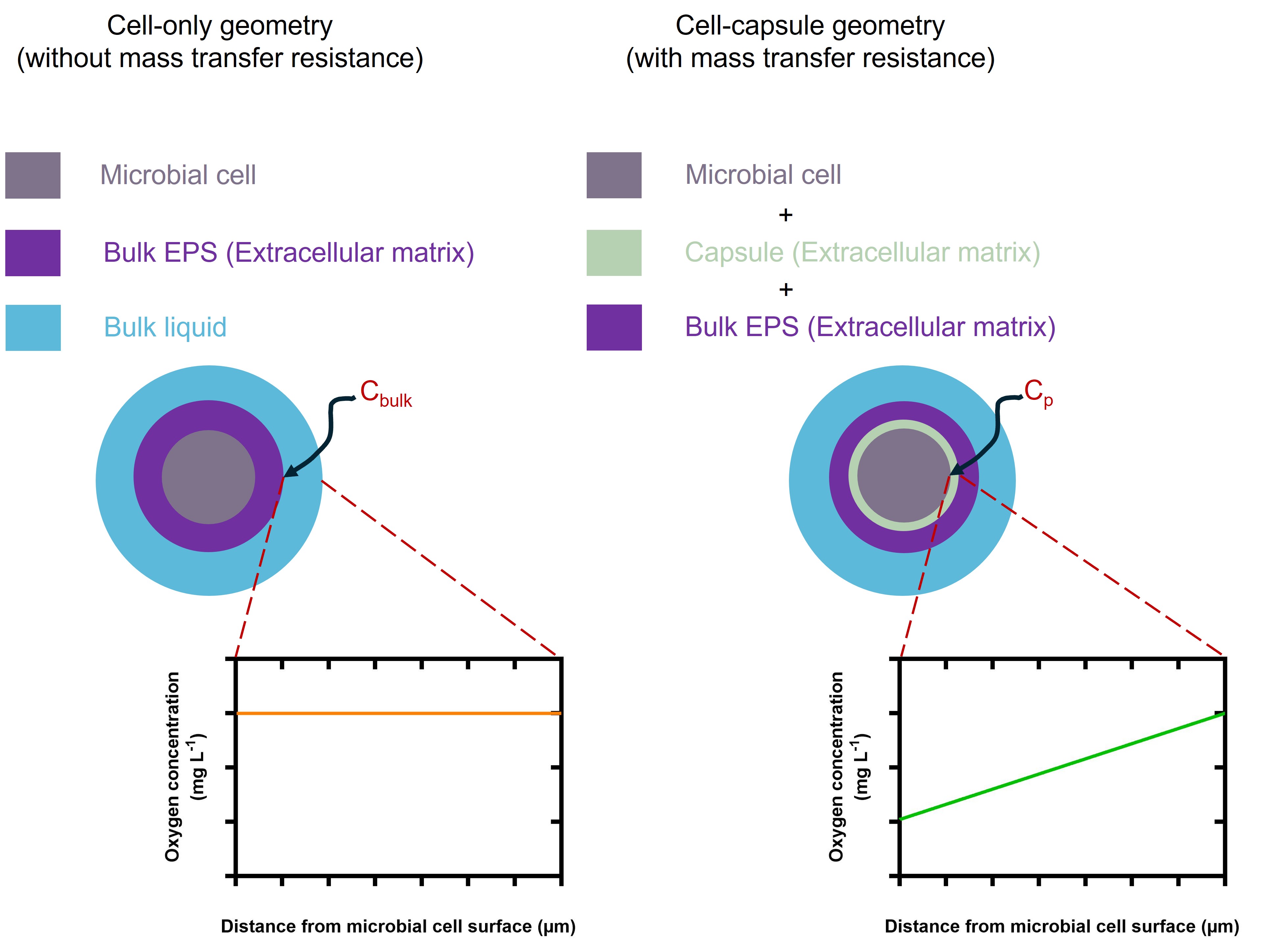}
		\centering
		\caption{\textbf{Role of capsule geometry of extracellular matrix as a 'resistance-in-series' model}: Representative spatial profiles of oxygen concentration subjected to with or without mass transfer resistance (model parameters are listed in Table \ref{tab:2D-model-without-flow-coupled-transport-parameters-list}) in the radial direction. A fixed capsule thickness ($\lpart$ = 0.70 $\mu$m) is assumed to simulate the mass transfer characteristics of the biofilm extracellular matrix. Note: Bacterial capsule and bulk EPS constitute the extracellular matrix around the microbial cell. Dimensions are not to scale in the above schematic diagram.}
		\label{2D-BPM-without-flow-MT-resistance-model-v5}
	\end{figure}

	Because the bacterial capsule has material properties distinct from the bulk extracellular matrix, our objective was to investigate this distinct difference in terms of oxygen transport. A number of models have conceptualized biofilm structures in which cells and the EPS matrix occupy distinct layers while exhibiting different oxygen diffusivities (\cite{Wanner2006}; \cite{Stewart1998}; \cite{Hulst1989a}; \cite{Cleary1979}). Therefore, the relative diffusional resistance of the capsule compared to the standard EPS matrix provided the foundational basis for our theoretical investigations. The model assumed that the resistance to mass transfer in the capsule is higher than that in the matrix, and the capsule thickness, $\lpart$, with typical dimensions in the range of hundreds of nanometers (\cite{Whitfield2020a}; \cite{Eichner2025a}), is a measure of this resistance. A simple reaction-diffusion analysis allowed prediction of the oxygen concentration profiles (see supplementary details, Figures S2 and S3). Our simulations distinctly illustrate the impact of the capsule through the aforementioned 'resistance-in-series’ model during the initial phases of biofilm development, where the steady-state oxygen concentration decreases by almost 70\% because of the elevated mass transfer resistance in the capsule surrounding the microbial cell (Figure \ref{2D-BPM-without-flow-MT-resistance-model-v5}).

	\subsection{Benchmarking standard mathematical model for oxygen uptake}
\begin{figure}[tbp]
		\includegraphics[width=1.0\textwidth]{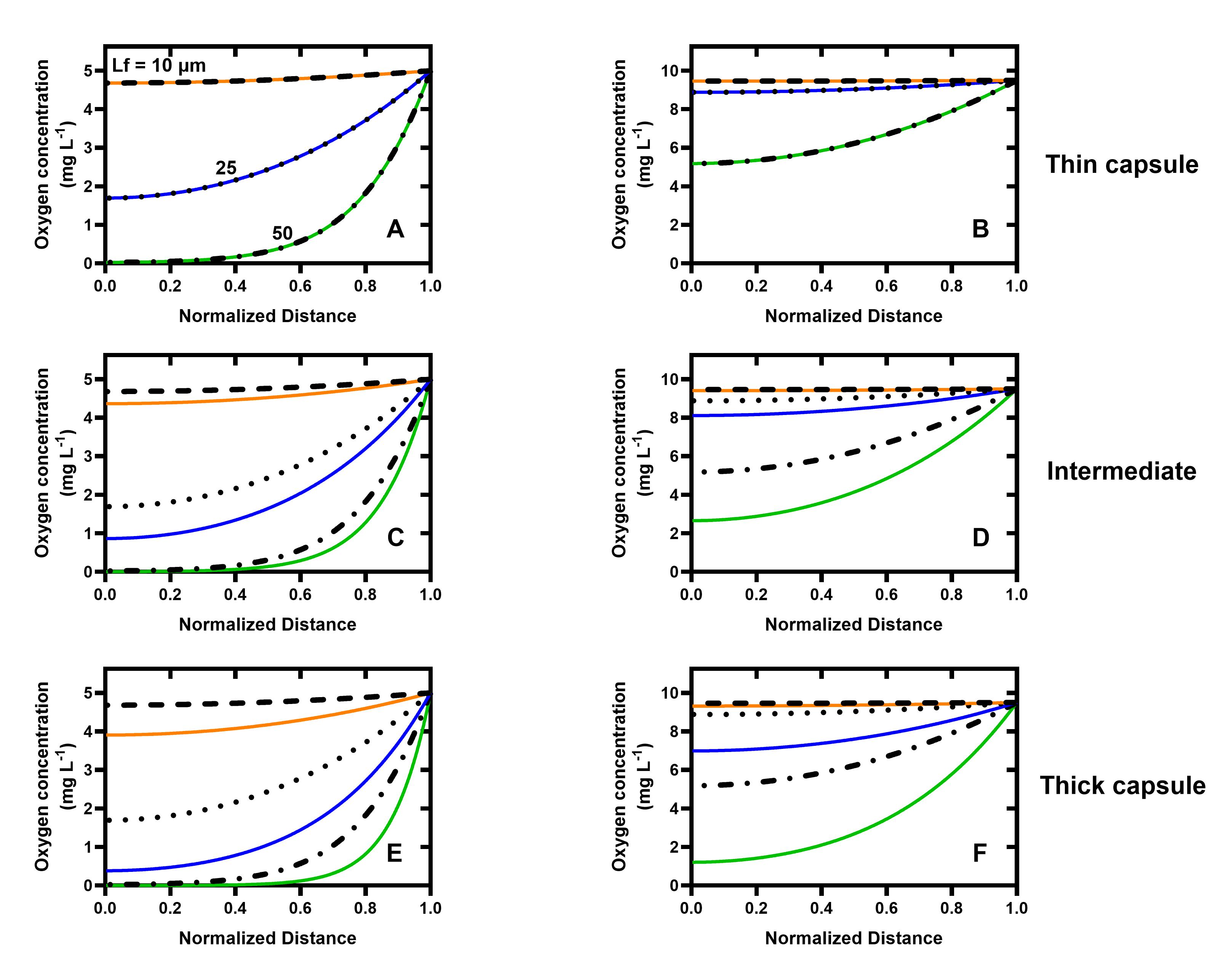}
		\centering
		\caption{\textbf{Benchmark analysis}: Comparison between standard reaction-diffusion model \citep{Stewart2016} and the present study for a known set of model kinetic parameters (refer to Table \ref{tab:2D-model-without-flow-coupled-transport-parameters-list-benchmark-analysis}) are shown using three different physiologically relevant thickness values ($\lpart$), and at two different initial oxygen concentrations ($\cbulkinitial$): (A) thin capsule ($\lpart$ = 0.14 $\mu$m), $\cbulkinitial$ = 5 mg L$^{-1}$ (B) thin capsule ($\lpart$ = 0.14 $\mu$m), $\cbulkinitial$ = 9.5 mg L$^{-1}$ (C) intermediate ($\lpart$ = 0.42 $\mu$m), $\cbulkinitial$ = 5 mg L$^{-1}$ (D) intermediate ($\lpart$ = 0.42 $\mu$m), $\cbulkinitial$ = 9.5 mg L$^{-1}$ (E) thick capsule ($\lpart$ = 0.70 $\mu$m), $\cbulkinitial$ = 5 mg L$^{-1}$ (F) thick capsule ($\lpart$ = 0.70 $\mu$m), $\cbulkinitial$ = 9.5 mg L$^{-1}$. The black-colored dotted lines (long-dashed, single-dotted or dot-dashed) represent the numerical solutions obtained from the standard model, and solid-colored lines (orange, blue or green) represent the numerical solutions obtained from our model for three different values of biofilm thickness ($\lf$ = 10, 25 or 50 $\mu$m) respectively. The model-based oxygen concentration profiles for the thin capsule thickness are observed to closely follow the standard model (see panels A and B). Note: Normalized distance is calculated as the ratio of distance from the substratum, $z$ to the biofilm thickness, $\lf$. Corresponding biofilm thickness is shown (see panel A) for the purpose of better visualization.}
		\label{fig:2D-BPM-without-flow-benchmark analysis-v11-with-standard-Monod-kinetics}
	\end{figure}

	Current models oversimplify the biofilm morphology (\cite{Characklis1978}; \cite{Wanner2006}), missing the complexity of the matrix and the cell arrangements or spacing. As a step towards a more structured model of the biofilm, we consider here the concept of a biofilm comprising of capsule surrounding the cells within a regular EPS matrix (see supplementary details, Figure S2). Our assumption that the capsule is denser than the bulk matrix is supported by previously published data (see supplementary details, Figure S4, Tables S1 to S4). These AFM studies demonstrate the capsule is significantly stiffer than the regular matrix, regardless of the capsule thickness. A recent study on reproducible measurements showed that the capsule varies in thickness from 0.1 to 1 µm \citep{Eichner2025a}. We chose to use three different capsule thicknesses in our simulations (Figure 2) representing thin, intermediate and thick capsule with values of  0.14 $\mu$m, 0.42 $\mu$m and 0.70 $\mu$m respectively. In the case of thick (0.70 $\mu$m) capsule, our model predicted relative deviations of more than 50\% from the standard model (panels E and F in Figure \ref{fig:2D-BPM-without-flow-benchmark analysis-v11-with-standard-Monod-kinetics}).

	\subsection{Effect of geometrical spacing on biofilm density}
\begin{figure}[tbp]
		\includegraphics[width=0.9\textwidth]{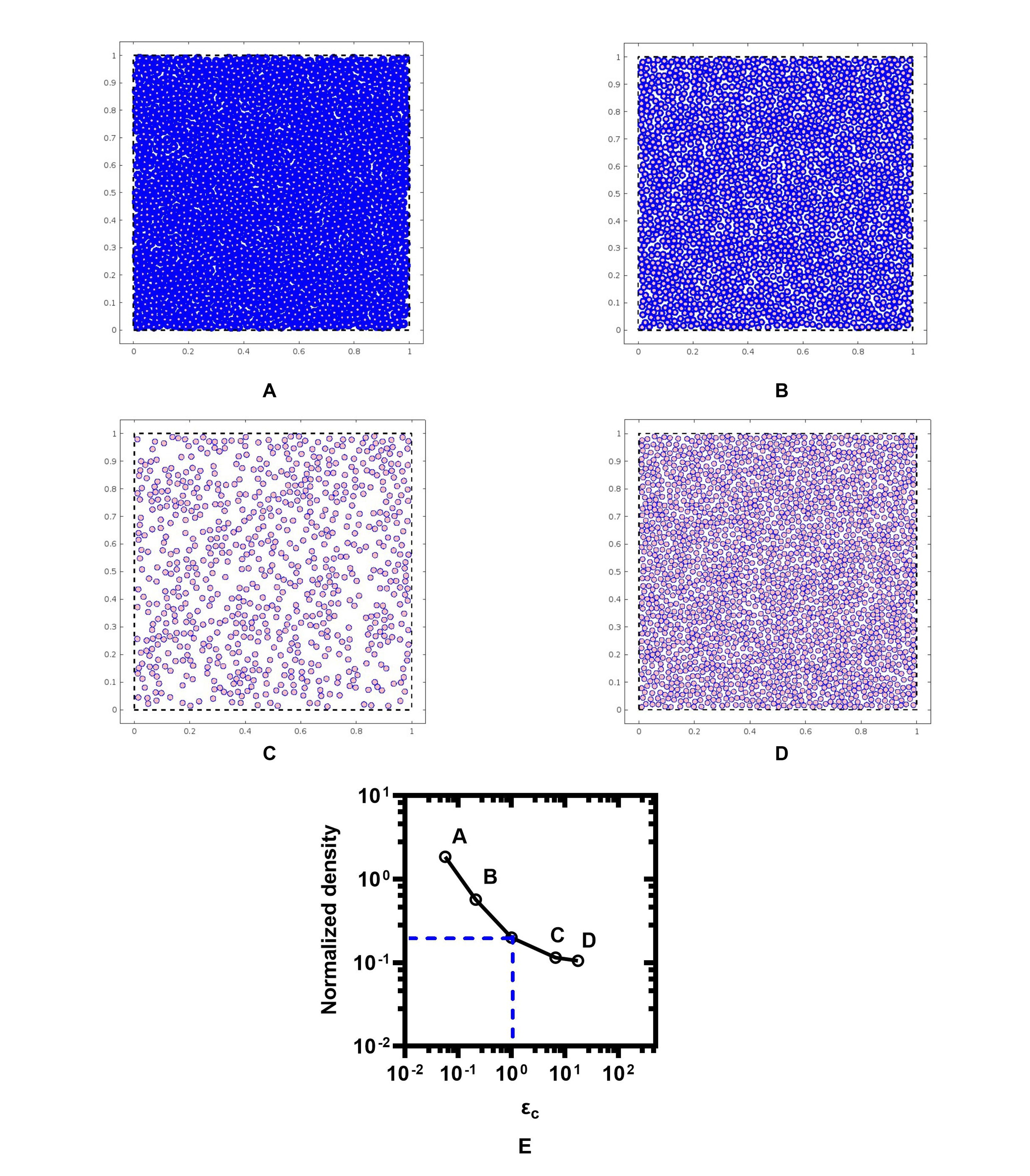}
		\centering
		\caption{\textbf{Effect of geometrical spacing on the density of extracellular matrix}: Typical cell-capsule patterns in the radial direction, x, generated as input datasets in our present study for a given, dimensionless unit surface area. For a fixed capsule thickness, $\lpart$ of 0.70 µm, different patterns are generated by varying the compactness factor ($\fpack$) as: (A) 0.057 $\pm$ 0.001 (B) 0.213 $\pm$ 0.008 (C) 6.677 $\pm$ 0.686 (D) 17.615 $\pm$ 1.104. Note: Random distribution of perfect spheres are subjected to image analysis using ImageJ software (ImageJ, NIH, USA), with blue-colored capsule around the pink-colored cells. (E) Graphical representation of different biofilm morphologies with varying compactness factor for cell-capsule patterns ($\fpack$) is plotted. The blue-colored dotted guide lines correspond to an estimate of normalized density at $\fpack$ = 1.002. A fixed capsule thickness, $\lpart$ = 0.70 $\mu$m is assumed for generation of the above spatial patterns, where $\fpack$ varies between 0.057 and 17.615. Note: Normalized density ($\rhoedr$) is estimated as the ratio of density of the capsule relative to the density of the EPS matrix. Trendline along the data points is shown for the purpose of better visualization.}
		\label{2D-BPM-without-flow-scenario-map-fpack-v5}
	\end{figure}

	Local density variations in the EPS matrix can lead to heterogeneous biofilm structures (\cite{Quan2022}; \cite{Geno2014a}; \cite{Dennis2018a}). Thicker capsule tends to form a dense EPS matrix, according to earlier experimental findings \citep{Phanphak2019a}. In this study, we investigate the relationship between capsule density and the compactness factor ($\fpack$). For example, a $\fpack$ of 0.057 corresponds to approximately 95\% of the biofilm space occupied by the cell-capsule patterns. To test different scenarios, we varied the geometrical spacing within the biofilm (or $\fpack$) from 0.05 to 50 (see Figure \ref{2D-BPM-without-flow-scenario-map-fpack-v5} for an example set of cell-capsule patterns). 
	\par Comparing the average density of the EPS matrix, which is roughly six times lower than the biomass density \citep{Horn2001}, we found that a highly compact matrix paired with a thick capsule remains relatively dense (Figure \ref{2D-BPM-without-flow-scenario-map-fpack-v5}). This suggests that the capsular structure may offer resistance to oxygen mass transfer around the cells, impacting the biofilm growth. Focusing on the capsule and spatial arrangements can thus be beneficial for designing future experiments to study heterogeneity in the matrix and for extracting compactness factors from advanced imaging of biofilms.
	
	\subsection{Mass transfer analysis using effectiveness factors in capsule region}
	Using the numerical reaction-diffusion analysis, effectiveness factors ($\etaparticle$) for bacterial capsule are examined with or without resistance to oxygen diffusion. By comparing these actual conditions to a standard scenario with no capsule thickness, we calculated effectiveness factors for different cell-capsule patterns. For example, an effectiveness factor of 1 is equivalent to no diffusion barrier to oxygen transport in the capsule (see methodology sections for more details).
	\par
\begin{figure}[tbp]
		\includegraphics[width=1.0\textwidth]{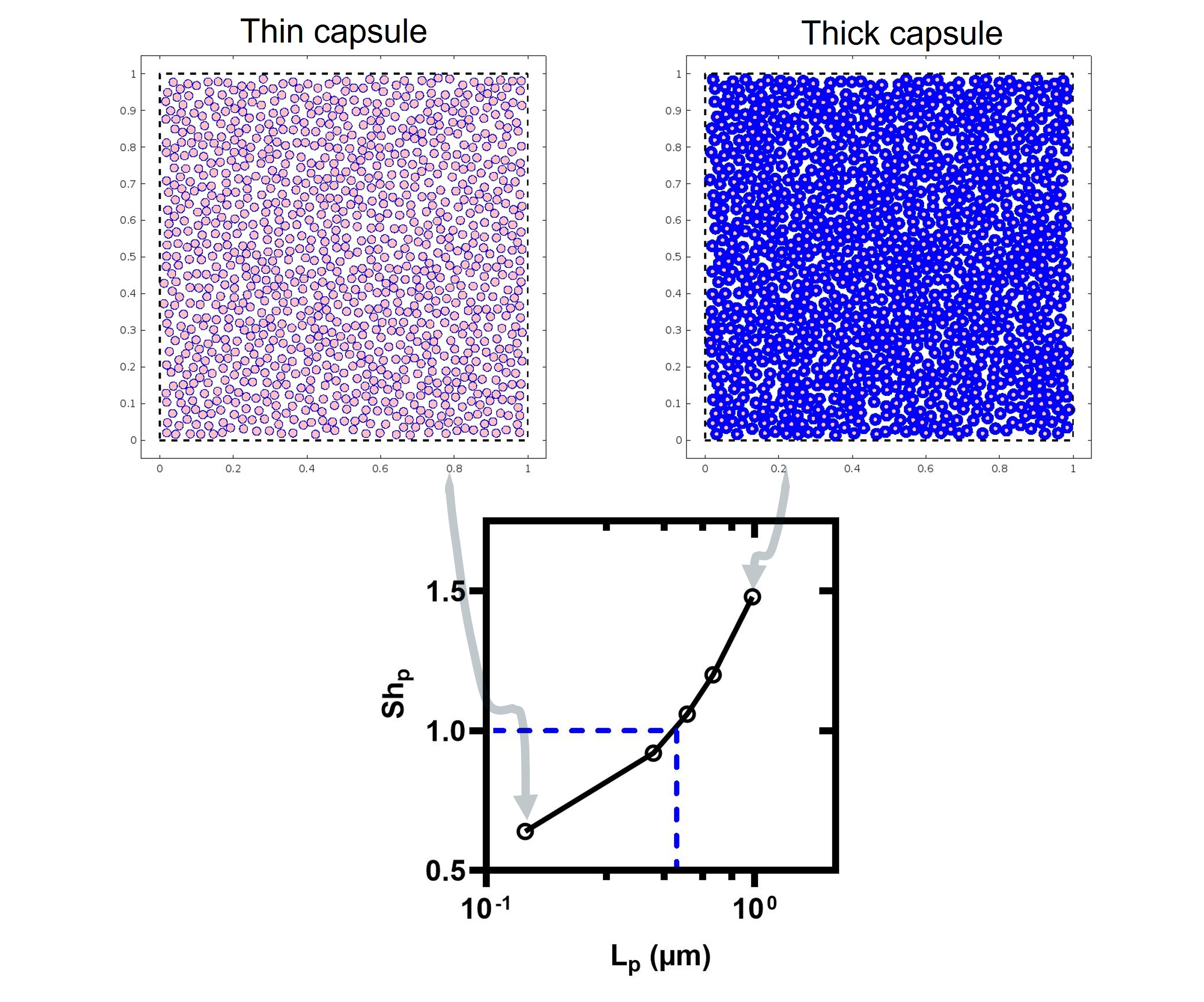}
		\centering
		\caption{\textbf{Effect of capsule thickness on the measure of mass transfer resistance due to extracellular matrix}: Graphical representation of different scenarios with varying capsule thickness ($\lpart$). For a given $\lpart$, Sherwood number ($\Shp$) is defined as the ratio of resistance due to oxygen diffusion and the resistance due to reaction kinetics for oxygen utilization in the capsule. Two representative cell-capsule patterns using thin capsule ($\lpart$ = 0.14 $\mu$m) and thick capsule ($\lpart$ = 0.98 $\mu$m) are shown. Note: Random distribution of perfect spheres are subjected to image analysis using ImageJ software (ImageJ, NIH, USA), with blue-colored capsule around the pink-colored cells. Trendline along the data points is shown for tracing the estimated variations in case of a given random distribution of spheres’ arrangement, and for the purpose of better visualization.}
		\label{2D-BPM-without-flow-scenario-map-v5}
	\end{figure}

 For a thick capsule ($\lpart$ = 0.7 $\mu$m or larger), we tested our hypothesis on the use of a structured model to analyze the diffusional limitations. These simulations predicted oxygen-limited conditions (Figure \ref{2D-BPM-without-flow-scenario-map-v5}), showing the importance of mass transfer resistance in the capsule. This is significant because the actual mass transfer condition is dependent on the value of the diffusion coefficient for oxygen in the capsule ($\dsparticle$). 
	\par
\begin{figure}[tbp]
		\includegraphics[width=1.0\textwidth]{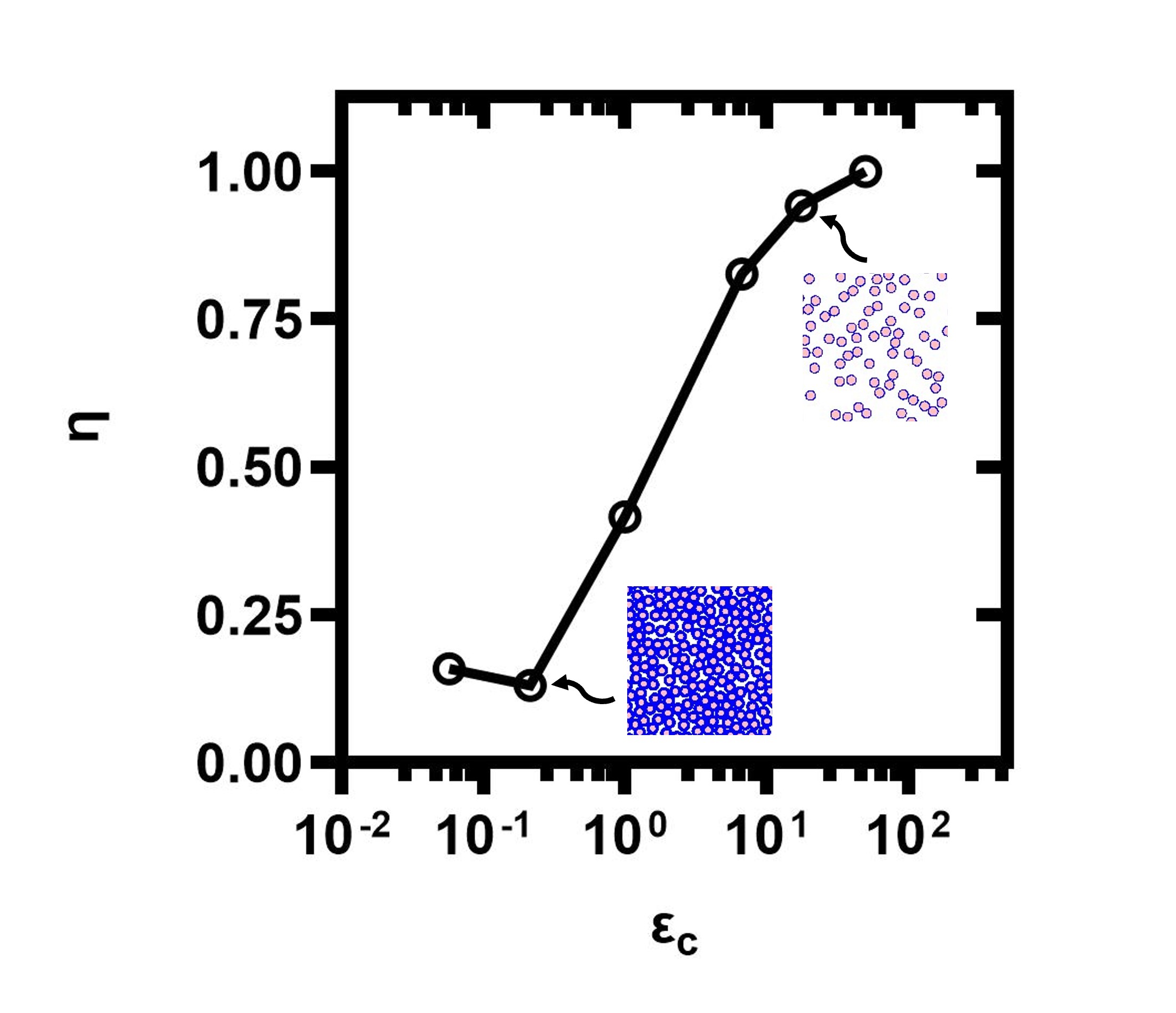}
		\centering
		\caption{\textbf{Predicted oxygen mass transfer performance in the capsule for various structured arrangements}: The curve shows effectiveness factors ($\eta$) obtained from oxygen profiles in the capsule region for varying compactness factor ($\fpack$). For densely arranged matrix (say, $\fpack$ of nearly 0.2 or lower), where diffusion resistance just begins to intrude, the effectiveness factor can slightly increase. The conceptual representation of the spatial patterns at corresponding effectiveness factors are provided. Note: Simulation results with different values of $\fpack$ are shown assuming a fixed capsule thickness, $\lpart$ = 0.70 $\mu$m.}
		\label{fig:2D-BPM-model-without-flow-MT-scaled-up-biofilm-profiles-v5}
	\end{figure}

 Considering a representative thickness of 0.7 $\mu$m, we investigate the mass transfer effectiveness of bacterial capsule using both loosely arranged and densely arranged matrices. We observed an interesting trend reversal as the effectiveness factor drops to around 0.2 or lower (Figure \ref{fig:2D-BPM-model-without-flow-MT-scaled-up-biofilm-profiles-v5}). Effectiveness factors near 0.1 clearly show cell-capsule patterns competing for space in the biofilm. Reduced effectiveness factors are close to 0.1 when the compaction factor falls to around 0.2 or less, proving that thick capsule can limit oxygen distribution, as cell-capsule patterns become denser. 
	\par Our model identifies limitations by adjusting geometric values, specifically the capsule thickness and compaction factor within the biofilm. We hereby demonstrate the dense matrix using different spatial patterns. Adhering to these structured arrangements, mass transfer in biofilms is likely to be diffusion-controlled. We used simulations to investigate the effect of bacterial capsule and its potential impact on cell organization and oxygen transport. In the case of thick capsule (\cite{Cleary1979}; \cite{Eichner2025a}; \cite{Khadka2025a}), our model demonstrates a significantly reduced impedance of oxygen transport from the biofilm liquid interface to the microbial cells, relative to conventional biofilm models that do not take into consideration the impact of capsule.  
	
	\subsection{Limitations of the model and future directions}
	While our multiscale framework successfully captures the impact of microscale heterogeneity on mass transport in structured biofilms, certain simplifying assumptions outline the scope and boundaries of the present work:
	\begin{enumerate}
		\item [a. ]
		\textit{Mass transport using poroelastic theory}: A primary consideration in our model is the derivation of diffusivity ratios $\left(\frac{\dsparticle}{\dbulk}\right)$ from experimental, AFM-based poroelastic metrics. While poroelasticity measures solvent migration rather than solute diffusion, it serves as a valid proxy for structural resistance since both processes are governed by the physical constraints posed by the same extracellular polymeric substance (EPS) volume fraction \citep{McCabe1975a}. This mechanical analogy is validated by the numerical alignment with independent \textit{in situ} oxygen uptake measurements ($\dsparticle$ $\approx$ 4 to 5*10$^{-5}$ m$^{2}$ day$^{-1}$) within dense capsular structures. This is consistent with early empirical observations demonstrating that dense capsular structures serve primarily as a metabolic barrier regulating cellular microenvironments \citep{Cleary1979}.
		\item [b. ]
		\textit{Modulus-diffusivity relationship}: Sensitivity analyses across a wide range of Young's moduli (6 to 320 kPa) demonstrate a flat diffusivity profile (see supplementary details, Figure S4). This plateau potentially identifies a saturation threshold supported by the specific polymer volume fraction, where the capsule’s internal structure becomes sufficiently dense \citep{Huang2026a} leading to matrix compaction. Consequently, further mechanical stiffening does not significantly reduce available free volume for oxygen transport. This suggests that the matrix compaction, rather than variations in absolute values of capsule thickness, is likely to be the primary driver of resistance to oxygen transport.
		\item [c. ]
		\textit{Range of oxygen diffusivity values}: The transport behavior simulated in our model relies on an established baseline of oxygen diffusivity in the bulk phase and its uptake within the capsule and the biofilm matrix. In reality, the local diffusivity of oxygen within a polysaccharide capsule is a dynamic variable governed by EPS hydrolysis, structural cross-linking, and localized metabolic consumption (\cite{Xavier2005c}; \cite{Whitfield2020a}; \cite{Gao2024a}; \cite{Cleary1979}). To ensure the metabolic parameters remain within a documented physiological range, we used biochemical saturation constants ($\khalfsat$) derived from the established biofilm kinetics literature (\cite{Rittmann1980}; \cite{Stewart2016}). However, direct \textit{in situ} measurements of oxygen gradients at the single-cell scale remain an open experimental challenge.
		\item [d. ]
		\textit{Bounded limits for capsular structures}: To maintain computational tractability and clearly demonstrate the transport-limiting signature of the capsule, we use a thin-shell approximation to determine the local diffusive flux, a common practice in mass-transfer limited microbial models, where core radius greatly exceeds the shell thickness \citep{Characklis1978}. Additionally, the structural parameter ($\kbiomassinactive$) was held constant to act as an intrinsic constitutive property of the model, ensuring that changes in the capsule's diffusion resistance are driven by the oxygen concentration gradients. 
		\item [e. ]
		\textit{Future experimental framework}: Another limitation of our work is the assumption of constant biomass density (\cite{Zhang2001a}; \cite{Eberl2001a}) throughout the biofilm and this can be addressed in future work. To refine these transport physics, we propose to relax this assumption to accommodate time-dependent density variations. In addition, the use of advanced imaging techniques (e.g., confocal microscopy at single-cell resolution (\cite{Dennis2018a}; \cite{Bottura2025a}) that distinguish capsule from bulk EPS is an experimental approach that will complement the present work. 	
	\end{enumerate}

	
	\section{Conclusion}
	This study introduces a novel approach to biofilm modeling that utilizes geometric patterns derived from explicit biofilm morphology. By evaluating distinct cell–capsule patterns, we provide a rigorous reaction–diffusion analysis that accounts for the intrinsic structural heterogeneity of the EPS matrix. Our conceptual "resistance-in-series" model successfully quantifies the transport impedance imposed by the bacterial capsule on local oxygen availability. Simulations demonstrate that capsule thickness and microscale geometric spacing dictate the transition into diffusion-controlled mass transfer regimes. Ultimately, this multiscale architecture framework offers an explainable mathematical foundation for predicting how microscale spatial organization regulates macroscale metabolic kinetics.

	\section{Supplementary information}
	The source code to generate spherical geometries is adapted from the MATLAB Help Center. The MATLAB codes used to numerically solve the one-dimensional governing equations and the worksheet on oxygen diffusivity estimates are available at the GitHub repository website of the ERC ABSOLUTE project here (https://github.com/raghukrm/ERC-ABSOLUTE-Biofilm-Models-BPM) (shared under CC-BY License).
	\section{Appendix}
	\label{AppendixEquations}
	\subsection{Governing equation for capsular region}
	The geometric domain of cell-capsule structure is approximated to the radial direction (or $x$-direction), with constant width ($\Wchannel$). It is defined by cell diameter ($\dparticle$) and capsule thickness ($\lpart$) (see supplementary details, Figure S2), and oxygen transport in the capsule geometry is given by the material balance including the diffusion and reaction components as follows:   \\
	\begin{equation}
		-\dsparticle \frac{\partial^2 \cparticle}{\partial x^2} - {\left(\frac{1}{\ybparticle}\right) \left(\frac{\kparticle \cparticle \rhoparticle}{\khalfsat + \cparticle}\right)} = 0
		\label{eqn:governing-PDE-biofilm-phase-particle}
	\end{equation} \\
	where, oxygen concentration in the capsule, $\cparticle$ varied along radial direction $x$, with diffusion coefficient of oxygen in the capsule, $\dsparticle\approx 0.2 \dbulk$ (assumed as 80\% less than that in the bulk phase, see supplementary details), and capsule density as $\rhoparticle$. In the capsular region, oxygen utilization is assumed with true-yield coefficient, $\ybparticle$, half-maximum rate concentration of oxygen, $\khalfsat$, and rate constant for oxygen utilization in the capsule, $\kparticle$ as reaction parameters. For a given rate decay coefficient for cell-capsule structures relative to changes in EPS matrix as $\kbiomassinactive$, and $\ninitialparticle$ cell-capsule structures initially arranged with a biofilm volume fraction of $\epart$; we calculated $\kparticle$ $\approx$ $\left[\left(\frac{\epart}{\ninitialparticle}\right) \left(\kbiomassinactive\right) \right]$. 
	
	\par An integration or a total sum of the cell-capsule structures along the radial direction ($x$) is re-scaled to estimate the oxygen concentration for a given biofilm thickness, say, $z$ (see supplementary details, Figure S2). Taking the basis of biofilm surface area and integrating the oxygen concentration over the given system dimensions (where, $\Wchannelradius = 0.5 \Wchannel$ by symmetry) in $x$ direction for every infinitesimal element area $\sim$ $\left(x \Delta x \right)$, we get,  
	\begin{equation}
		\nparticle = \left[\frac{\ninitialparticle}{\rhoparticle}\right]  \left[\int_0^\Wchannelradius \left(\frac{\ybparticle}{\Vbiofilmparticleeff}\right) \cparticle x \Delta x\right]
		\label{eqn:particle-density-function-governing-eqn}
	\end{equation} 
	where, $\nparticle$ is apparent number density based on cell-capsule patterns, $\dparticle$ is cell diameter, $\Vbiofilmparticleeff$ is effective volume of capsule and equivalent to:
	\begin{equation}
		\Vbiofilmparticleeff =\text{ }\pi \left(\left(\dparticle+\lpart\right)^2-\left(\dparticle\right)^2\right) \lpart
	\end{equation} 
	\subsection{Governing equation for biofilm phase}
	The geometry domain is approximated to the lateral direction (or $z$-direction), with constant width ($\Wchannel$), biofilm thickness, $\lf$ and surface area, $\Abiofilm \left(=\Wchannel\lf\right)$. Diffusive transport of oxygen and biomass growth kinetics due to oxygen utilization are included in the material balance as follows: \\
	\begin{equation}
		-\ds \frac{\partial^2 \cs}{\partial z^2} - {\left(\frac{1}{\yb}\right) \left(\frac{\cs}{\khalfsat + \cs}\right) \Abiofilm\freqpart \rhoxb} = 0
		\label{eqn:governing-PDE-biofilm-phase}
	\end{equation} \\
	where, oxygen concentration in biofilm phase, $\cs$ varied along lateral direction $z$, with diffusion coefficient of oxygen in the biofilm, $\ds$ (20\% less than that in the bulk phase, or $\approx 0.8 \dbulk$) \cite{Stewart1998}, and biomass density as $\rhoxb$. The growth of biomass is assumed to follow Monod kinetics due to oxygen utilization with yield coefficient, $\yb$, half-maximum rate concentration of oxygen, $\khalfsat$, and rate of particulate density based on cell-capsule patterns, $\freqpart$ as the reaction parameters. \\
	\begin{equation}
		\freqpart = \constone \left(1+\left(\frac{\nparticle+\consttwo}{\constthree}\right)\right)^{-\constfour}
		\label{eqn:Pareto-distribution}
	\end{equation} \\
	where, the biomass growth rate in the biofilm represented by $\freqpart$ is modelled as a function of the apparent number density based on cell-capsule patterns, $\nparticle$ and a set of four distribution parameters, namely $\constone$, $\consttwo$, $\constthree$ and $\constfour$, as derived from the cell-capsule structures (see Equation \ref{eqn:particulates-PSD-general-parameters} and Table \ref{tab:2D-model-without-flow-coupled-transport-parameters-list}). 
	
	\subsection{Governing equation for bulk phase}
	Assuming stationary bulk-biofilm interface, oxygen transport in the bulk liquid is given by the material balance including the diffusive and reaction components as follows (Equation \ref{eqn:governing-PDE-bulk-phase}):
	\begin{equation}
		-\dbulk \frac{\partial^2 \cbulk}{\partial z^2} - \aspbulk \kf\left(\cbulk - \frac{\left(1-\ebulk\right)}{\ebulk}\cs\right) = 0
		\label{eqn:governing-PDE-bulk-phase}
	\end{equation}  
	where, oxygen concentration in bulk phase, $\cbulk$ varied along $z$-direction, with diffusion coefficient of oxygen in the bulk phase as $\dbulk$, and varying surface area as $\aspbulk \left(=\frac{\Wchannel+\left(\Hchannel-z\right)}{\Wchannel \left(\Hchannel-z\right)}\right)$. An external mass transfer coefficient at bulk-biofilm interface, $\kf$ is estimated from previous literature \cite{Comiti2000} to apply a linear driving force with reference to the corresponding oxygen concentration obtained from the biofilm phase, $\cs$.

	\subsection{Definition of domain boundaries}
	Our model assumes the microbial cell (as an unreacted core geometry) with a constant oxygen concentration at its surface, or in close contact with the capsule. This condition is imposed in the radial direction at $x = 0.5$ $\dparticle$, and as given by Equation \ref{eqn:biofilm-phase-BC-part-derived}: 
	
	\begin{equation}
		{\left(\cparticle\right)}_{x} = \Shp \cbulkinitial
		\label{eqn:biofilm-phase-BC-part-derived}
	\end{equation} 
	where, $\cparticle$ is the oxygen concentration in the capsule, $\cbulkinitial$ is the initial oxygen concentration, and $\Shp$ is the dimensionless Sherwood number for the cell-capsule structure, and as obtained from Equation \ref{eqn:particle-Sherwood-number-formulation}.
	
	\par Oxygen transfer is assumed zero at the substratum. The oxygen concentration gradient is set to zero at one of the system boundaries at $z = 0$ in our model, and as given by Equation \ref{eqn:biofilm-phase-BC}:  
	\begin{equation}
		\left(\frac{\partial \cs}{\partial z}\right)_{z} = 0
		\label{eqn:biofilm-phase-BC}
	\end{equation}
	
	\par Diffusive transport in bulk liquid is considered to maintain a constant supply of oxygen-rich conditions at steady state. The initial oxygen concentration, $\cbulkinitial$ is assumed to be available at the bulk-biofilm interface (say, $z = \lf$) in the bulk phase during the biofilm development as follows (Equation \ref{eqn:bulk-phase-BC-BPM}):
	\begin{equation}
		\left({\cbulk}\right)_{z} = \cbulkinitial
		\label{eqn:bulk-phase-BC-BPM}
	\end{equation}
	where, $\cbulk$ is oxygen concentration in bulk phase and $\lf$ is the biofilm thickness.

	\section{Declaration of Competing Interest}
	All co-authors hereby state that there is no conflict of interest towards their contribution in the present work. 

	\section{Declaration of Generative AI and AI-assisted technologies in the writing process}
	During the preparation of this work the author(s) used Grammarly Pro in order to improve the readability and the writing style. After using this tool, the author(s) reviewed and edited the content as needed and take(s) full responsibility for the content of the publication.
	
	\section{CRediT author statement}
	Raghu K. Moorthy: Conceptualization, methodology, validation, formal analysis, investigation, data curation, writing - original draft, visualization. Eoin Casey: Conceptualization, methodology, formal analysis, resources, writing - review \& editing, visualization, supervision, project administration, funding acquisition.
	\section{Acknowledgements}
	We are thankful to the computational facilities provided at School of Chemical and Bioprocess Engineering, University College Dublin (UCD). Authors (RKM and EC) duly acknowledge funding received through ERC Advanced Grant for ABSOLUTE project, funded by European Research Council (ERC) (Grant No. 101052376), to support our work at University College Dublin (UCD). Our discussions within the ABSOLUTE project team (Dr. Dishon, Jacobus, Ciaran and Dibya) has been helpful in developing the model parameters.
	\section{Data availability}
	All datasets generated using MATLAB code in this present work are available on our project GitHub repository website of the ERC ABSOLUTE project here (https://github.com/raghukrm/ERC-ABSOLUTE-Biofilm-Models-BPM) (shared under CC-BY License).
	\printnomenclature[2.5cm]
	
	\bibliography{mybibfile}

\end{document}